\documentclass[aps,amsmath,amssymb,reprint,twocolumn,prl,superscriptaddress,notitlepage,showpacs,tightenlines]{revtex4-1}

\usepackage{exscale}
\usepackage{relsize}
\usepackage{graphicx}
\usepackage{dcolumn}
\usepackage{bm}
\usepackage{textcomp}
\usepackage[utf8]{inputenc}
\usepackage[T1]{fontenc}
\usepackage{mathptmx}
\usepackage{siunitx}
\usepackage{appendix}
\usepackage{array}
\usepackage{color}
\usepackage{textcomp}
\usepackage{amsmath}
\usepackage{float}
\usepackage{verbatim}
\usepackage[colorlinks,citecolor=blue,linkcolor=red,hyperindex,CJKbookmarks]{hyperref}

\allowdisplaybreaks[2]

\begin{document}
\title{Hybrid quantum surface acoustic wave with skyrmion qubit for quantum information processing}

\author{Yu-Yuan Chen}
\affiliation{School of Integrated Circuits, Tsinghua University, Beijing 100084, China}
\affiliation{Frontier Science Center for Quantum Information, Beijing 100084, China}

\author{Zhihui Peng}
\affiliation{Key Laboratory of Low-Dimensional Quantum Structures and Quantum Control of Ministry of Education, Department of Physics and Synergetic Innovation Center of Quantum Effects and Applications, Hunan Normal University, Changsha 410081, China}

\author{Yu-xi Liu}
\email{yuxiliu@mail.tsinghua.edu.cn}
\affiliation{School of Integrated Circuits, Tsinghua University, Beijing 100084, China}
\affiliation{Frontier Science Center for Quantum Information, Beijing 100084, China}
\date{\today}

\begin{abstract}
Surface acoustic wave (SAW) devices are key components of classical communication systems and recently studied for quantum information processing. We here propose and study a hybrid quantum system composed of skyrmion qubit and a SAW cavity, which supports a number of long-lived phononic modes. The results show that the system allows for strong coupling between skyrmion qubit and single phonon of different modes. By manipulating the qubit(s) through a static magnetic field and a time-dependent modulation magnetic field, we further study the interaction between skyrmion qubit and individual phononic modes, phonon-phonon interaction, and qubit-qubit interaction, which operates in strong-coupling regime. The controllability of nanoscale skyrmion qubit and the dense phononic modes of single SAW cavity would make our system have promising applications in large-scale quantum communication and computing.
\end{abstract}

\maketitle

%%%%%%%%%%%%%%%%%%%%%%%%%%%%%%%%%%%%%%%%%%%%%%%%%%%%%%%%%%%%%%
{\it Introduction.---}
\label{Introduction}
With the rapid development of the quantum information processing, superconducting qubit circuits have been considered as promising solid-state quantum computing platforms, due to their advantages in design flexibility,  tunability, scalability and interface capability with other quantum systems~\cite{Superconducting-Circuit1,Superconducting-Circuit2,Superconducting-Circuit3}. However, the superconducting qubits have shortages of relatively short coherent time and non-identity in contrast to natural spin systems.  By using the advantages of different systems,  hybrid quantum systems consisting of superconducting circuits and other systems (e.g., atoms, ions, and spins) have attracted extensive attentions~\cite{Superconducting-Circuit1,Superconducting-Circuit2}.  Advances in the hybrid quantum systems have shown great potential in quantum information processing~\cite{Superconducting-Hybrid,Superconducting-Hybrid-Chip}.

The miniaturization of hybrid microwave devices~\cite{Cavity-Bus1,Cavity-Bus2, Cavity-Register1,Cavity-Register2}  has stimulated great interest in integrating superconducting qubits with quantized surface acoustic waves (SAWs) ~\cite{QuantumAcoustics-Multimode-AW,QuantumAcoustics-Waveguide,QuantumAcoustics-Transducer1-Article1-110-PESAW-Quantization,
QuantumAcoustics-Article2,QuantumAcoustics-SAW-Superconducting-Coupling1,QuantumAcoustics-SAW-Superconducting-Coupling2,SAW-Sensing,
SAW-Entanglement1,SAW-Entanglement2,SAW-Piezomagnetism}, which are the mechanical waves propagating in elastic medium and have about $10^5$ times reduction in propagation velocity as compared with electromagnetic waves~\cite{SAW-Background-Book1,SAW-Background-Book2,SAW-Sensor,SAW-Fluid-Control}. That means, the wavelength of SAW is about $10^5$ times smaller than that of electromagnetic wave at the same frequency. These enable SAW devices (including cavities or waveguides) to link different quantum components in more compact way compared with the microwave transmission line cavities or waveguides, and thus have potential applications in quantum information processing as for modern microelectronic communication~\cite{SAW-Background-Book1,SAW-Background-Book2}. Moreover, quantized SAW devices can support a number of long-lived phononic modes~\cite{SAW-Cavity,QuantumAcoustics-Multimode-SAW,SAW-Cavity-Multimode-TunableCoupling}. Therefore,  quantized SAW devices are very suitable for quantum data bus or register.

Spins are conventional systems for quantum information storage due to their long coherent time and small size. However, natural spins have difficulty in fast individual manipulation and flexible design of system parameters. Skyrmions, the smallest possible configurations for an ensemble of magnetic spins, have localized nanoscale structures and exhibit topologically stable quasi-particle properties~\cite{Skyrmion-First-Theory,Skyrmion-Review1,Skyrmion-Review2,Skyrmion-Review-Devices1,Skyrmion-Review-Devices2,Skyrmion-Review-Devices3}. Thus, skyrmions are expected to be used as future data-storage or information carriers for long-lived spintronic devices~\cite{Skyrmion-Review-Devices1,Skyrmion-Review-Devices2,Skyrmion-Review-Devices3}. Most recently, Skyrmions have emerged as promising qubits for the storage of quantum information~\cite{SkyrmionQubit-First-Theory,SkyrmionQubit-First-Theory-Helicity,SkyrmionQubit-First-Review,SkyrmionQubit-QuantumComputing,SkyrmionQubit-Magnon}. To develop miniaturized platform for separately processing and storing quantum information by superconducting and skyrmion qubits,  it is necessary to explore a data bus for compactly connecting skyrmion to superconducting qubits.

In view of the achievements for the strong coupling between superconducting qubits and SAW phonons~\cite{SAW-Cavity-Multimode-TunableCoupling}, we here study a hybrid quantum system composed of skyrmion qubit and a SAW cavity supporting a number of phononic modes. We find that the coupling strength between single skyrmion qubit and single phonon of different modes can be in strong coupling regime. Based on this, we further study the interaction between skyrmion qubit and individual phononic modes, phonon-phonon interaction, and qubit-qubit interaction, which are essential in quantum information storage. The results show that these interactions also operate in strong-coupling regime. Therefore, our study provides a highly attractive platform for quantum information processing.

\begin{figure}[tb]
	\centering
	\includegraphics[width=8.6cm]{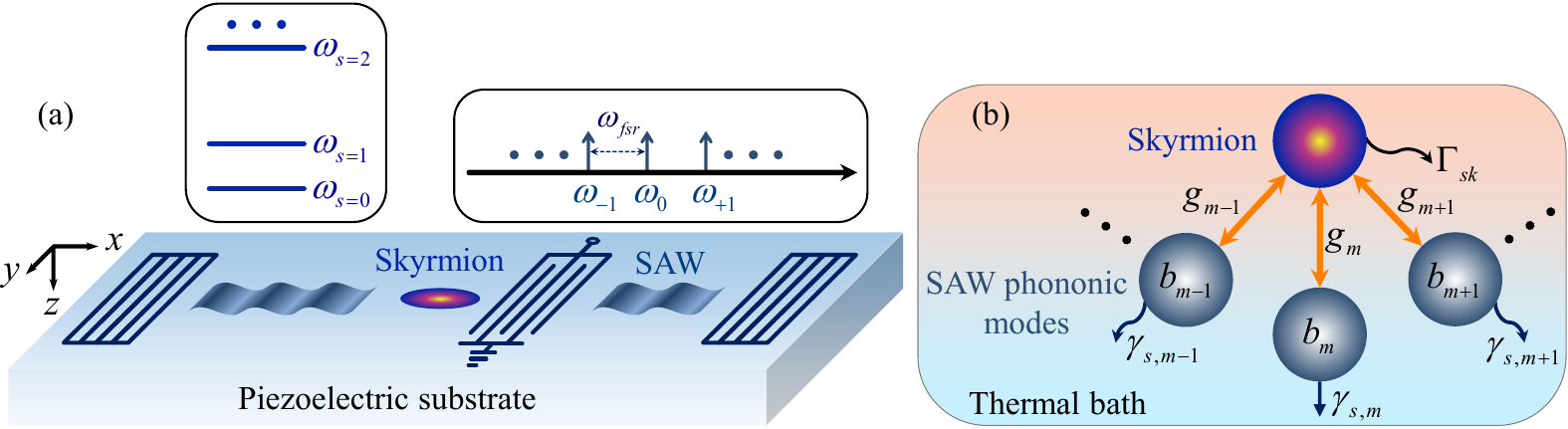}
	\caption{(a) Schematic illustration for coupling single skyrmion qubit to phonons in single SAW cavity via the electric field induced by piezoelectric effect. Insets show the energy-level structure of qubit and the frequency space diagram of phononic modes. (b)Diagram for the interaction between single skyrmion qubit (denoted by purple ball) and SAW pnonons of different modes (denoted by gray balls) under a common thermal bath.}
\label{F1System}
\end{figure}

%%%%%%%%%%%%%%%%%%%%%%%%%%%%%%%%%%%%%%%%%%%%%%%%%%%%%%%%%%%%%%
{\it Theoretical model.---}
\label{System}
As schematically shown in Fig.~\ref{F1System}(a), we consider a hybrid quantum system composed of a skyrmion and a quantized multimode SAW cavity deposited on the surface of piezoelectric substrate. The skyrmion is coupled to phononic cavity via electric field induced by piezoelectric effect. The Hamiltonian is written as~\cite{SkyrmionQubit-First-Theory,Skyrmion-Competing-Interactions,MagneticSpins-Electric-Control,Supplemental-Material}
\begin{equation}
\tilde{H}=
\overline{S} \int
d\bm{r}\left[
\frac{K_z}{a^2} m_z^2-
\frac{H_z}{a^2} m_z-
\frac{H_{\perp x}}{a^2} x m_x-
\bm{E} \cdot \bm{P}
\right]+
H_{saw},
\label{eq:1}
\end{equation}
where $\overline{S}$ denotes the magnitude of the effective spin. $K_z$ is the anisotropy coupling strength. $H_z$ is the static magnetic field strength in $z$-direction. $H_{\perp x}$ is the magnetic field gradient in $x$-direction. $m_z$ and $m_x$ are magnetizations along $z$- and $x$-directions. $a$ is the lattice space. The first three terms describe the skyrmion, coupled with the magnetic fields $H_z$ and $H_{\perp x}$. The fourth term describes the interaction between the skyrmion and the electric field~(see detail in supplementary materials~\cite{Supplemental-Material}) $\bm{E}=a \bm{e}_z \sum_m E_m^{zp} \text{sin} (k_m x) \left(b_m +b_m^{\dagger}\right)$ produced by quantized SAW. $\bm{e}_z$ is the normalized vector in $z$-direction. $E_m^{zp}$ denotes the zero-point fluctuation of electric field induced by single phonon with frequency $\omega_m$. $k_m =\omega_m /v_s$ is the wavenumber depending on the propagation velocity $v_s$ of SAW. $b_m^{\dagger}$ ($b_m$) is the creation (annihilation) operator of SAW phonon with frequency $\omega_m$. $\bm{P}$ is the electric polarization due to a pair of non-parallel spins. The last term $H_{saw}=\sum_m \omega_m b_m^{\dagger} b_m$ denotes the multimode SAW phonons. By projecting the magnetic skyrmion to its two lowest quantized energy levels of the deviation from equilibrium~\cite{Supplemental-Material}, we have the full Hamiltonian
\begin{align}
H&
=\omega_q \sigma_{11}
+\sum_{m} \omega_m b_m^{\dagger} b_m
+\sum_{m} g_m \big(b_m \sigma_{10} +\text{H.c.}\big)
\label{EQ1Hamiltonian}
\end{align}
under the rotating-wave approximation. Hereafter we take $\hbar=1$ for simplicity. The skyrmion is assumed to have high anharmonicity larger than $20\%$~\cite{Supplemental-Material}, such that we can isolate two lowest-energy levels as qubit. Due to the narrow free spectral range of SAW cavity, the qubit interacts with several phononic modes. Here, $\omega_q$ is the transition frequency between the ground $|0\rangle$ and excited $|1\rangle$ states of the skyrmion, which can be controlled by a static magnetic field in $z$ direction. $\sigma_{ij}=|i\rangle\langle j|$ is the Pauli operator of skyrmion, with $i,j=0,1$. $g_m=g_{m,0} \text{sin}(k_m x)$ denotes the coupling strength between the qubit at position $x$ and single phonon of frequency $\omega_m$, with the amplitude $g_{m,0}$ determined by the zero-point fluctuation $E_m^{zp}$ of electric field and skyrmion intrinsic properties~\cite{Supplemental-Material}. The quantum Langevin equations of the system are given as
\begin{align}
\frac{d \hat{b}_m}{dt}&
=-i {\omega}_m^e \hat{b}_m-i g_m \hat{\sigma}_{01}+\hat{F}_m,
\nonumber\\
\frac{d \hat{\sigma}_{01}}{dt}&
=-i {\omega}_q^e \hat{\sigma}_{01}
-i \sum_m g_m \hat{b}_m\left(\hat{\sigma}_{11}-\hat{\sigma}_{00}\right)+\hat{F}_{01},
\label{EQ2LangevinEqs}
\end{align}
with ${\omega}_m^e={\omega}_m -i\gamma_{s,m}\left(n_{th,m}+1\right)/2$ and ${\omega}_q^e={\omega}_q -i \Gamma_{s k}\left(n_{th,sk}+1\right)/2$. $\Gamma_{sk}$ denotes the decay rate of the skyrmion qubit. $\gamma_{s,m}=\gamma_{s,t} +\gamma_{s,d}(\omega_m) +\gamma_{s,i}(\omega_m)$ is the decay rate of phononic mode $\omega_m$, with the three terms arising from the grating transmission, grating diffraction, and coupling with interdigital transducer (IDT), respectively~\cite{Supplemental-Material}. $n_{th,m}=1\big/\left(e^{\hbar \omega_m / k_B T}-1\right)$ and $n_{th,sk}=1\big/\left(e^{\hbar \omega_q / k_B T}-1\right)$ are thermal populations for phononic modes and qubit at the temperature $T$. $\hat{F}_m$ and $\hat{F}_{01}$ are the quantum fluctuations corresponding to variables $\hat{b}_m$ and $\hat{\sigma}_{01}$.

%%%%%%%%%%%%%%%%%
%%%%%%%%%%%%%%%%%

%%%%%%%%%%%%%%%%%
\begin{figure}[tb]
	\centering
	\includegraphics[width=8.6cm]{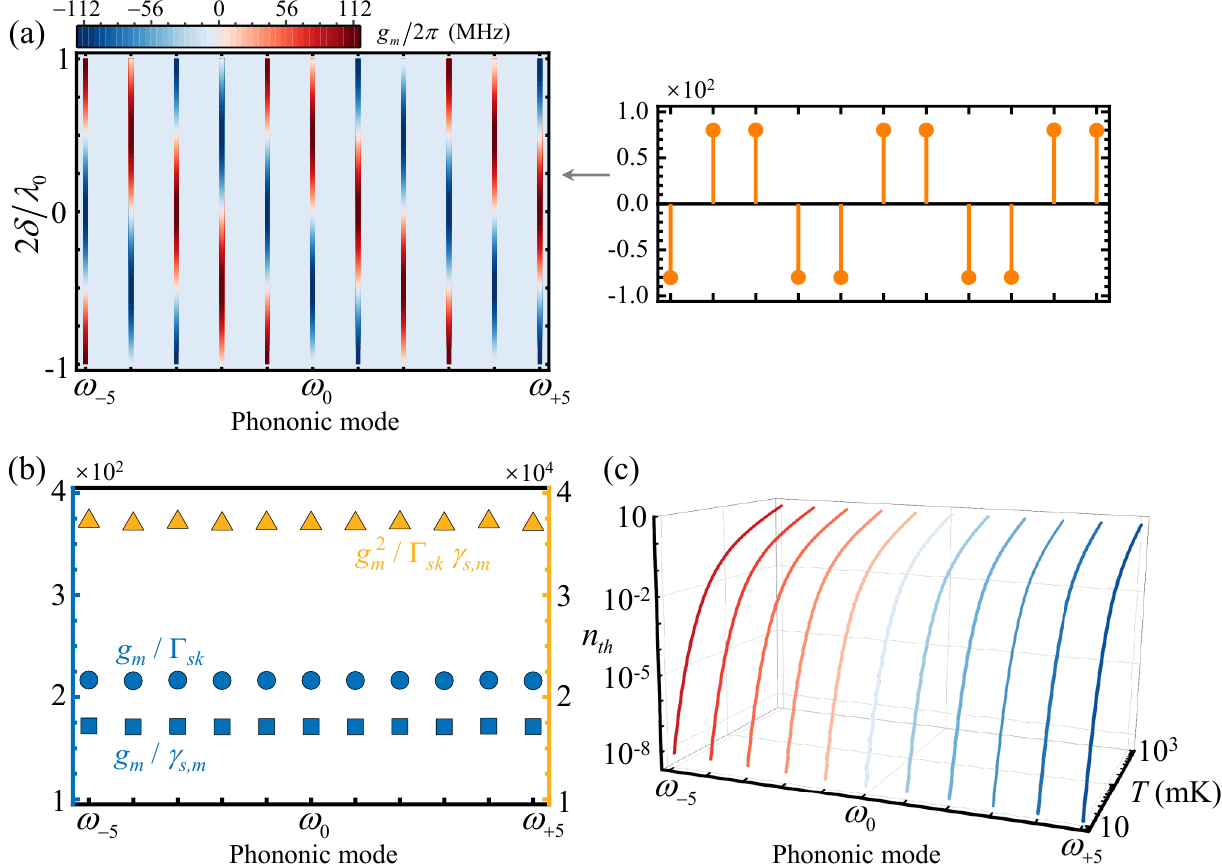}
	\caption{(a) Coupling strengths $g_{m}$ between a skyrmion qubit and single phonon of different modes for different skyrmion position $x=L_c/2+\delta$. Inset shows the details for $\delta=\lambda_0/8$. (b) Coupling-decay ratios $g_m / \Gamma_{sk}$ and $g_m / \gamma_{s,m}$ along with cooperativity $g_m^2 \big/ \Gamma_{sk} \gamma_{s,m}$ for different phononic modes when taking $\delta=\lambda_0 /8$. (c) Thermal population for different phononic modes versus bath temperature $T$. The parameters are $g_{m,0}/2\pi =113\,\text{MHz}$, $\omega_q /2\pi \simeq \omega_0 /2\pi =4\,\text{GHz}$, $\Gamma_{sk}/2\pi=0.4\,\text{MHz}$, $\gamma_{s,t}/2\pi=6.2\,\text{kHz}$, $\sqrt{\omega_m \cdot \gamma_{s,d}}/2\pi=1.42\,\text{MHz}$, and $\gamma_{s,i}/2\pi=0.5 \text{sinc}^2 \left[10 \pi ({\omega_m -\omega_0})/{\omega_0}\right]\,\text{MHz}$.}
\label{F2StrongCoupling}
\end{figure}
%%%%%%%%%%%%%%%%%

{\it Strong coupling.---}
Strong couplings between the skyrmion qubit and single phonon of different modes are necessary for quantum coherent operations. As shown in the inset of Fig.~\ref{F1System}(a), we assume that the SAW cavity supports several phononic modes with center frequency $\omega_0 /2\pi=4\,\text{GHz}$ and wavelength $\lambda_0=v_s/\omega_0$, accompanied by neighbouring modes with frequency $\omega_m=\omega_0 + m \omega_{fsr}$~(with mode index $m=\pm1,\pm2...$). $\omega_{fsr}=v_s \big/ 2L_c$ is the free spectral range, determined by the effective cavity length $L_c$. As illustrated in Fig.~\ref{F2StrongCoupling}(a), due to the distinct node locations of different standing-wave phononic modes, the coupling strengths between the skyrmion qubit and phonons of different modes highly depend on the position of the skyrmion. Furthermore, we assume that the qubit with the transition frequency $\omega_q \simeq \omega_0$ is at the position $x$ with a deviation $\delta$ from the center of the cavity, i.e., $x=\delta +L_c/2$. Thus, by adjusting the position of skyrmion~(typically with current, electric or magnetic field~\cite{Skyrmion-Review-Devices3,Skyrmion-Review1,Skyrmion-Motion-Control1,Skyrmion-Motion-Control2}), the qubit can be coupled to phonons of specific phononic modes. For example, by taking $\delta=0$ (or $\delta=\lambda_0/2$), the qubit is coupled to only the phonons of modes $m=0,\pm2,\pm4$ (or $m=\pm1,\pm3,\pm5$). Specifically, by taking $\delta=\lambda_0/8$, the amplitudes $|g_m|$ of coupling strength are approximately equal for different phononic modes. In such a case, as shown in Fig.~\ref{F2StrongCoupling}(b), the coupling strength between the qubit and single phonon of different modes is much larger than the decay rates of qubit and phonons (i.e.,~$|g_m|\gg\{\Gamma_{sk},\,\gamma_{s,m}\}$), accompanied by high cooperativity (i.e., $g_m^2 \big/ \Gamma_{sk} \gamma_{s,m} \simeq 3.7\times10^4$). Thus,  the skyrmion-SAW hybrid quantum system can operate in strong-coupling regime.

When the present system is coupled to a bath with a finite temperature $T$ as illustrated in Fig.~\ref{F1System}(b), the thermal populations $n_{th,m}$ and $n_{th,sk}$ would result in the increase of decay rates as indicated in Eq.~(\ref{EQ2LangevinEqs}). This may destroy the multimode strong-coupling regime, which is necessary for coherent operation. Figure~\ref{F2StrongCoupling}(c) gives the thermal population for different phononic modes under different temperature $T$. Note that for the bath temperature $T<100\,\text{mK}$, the effect of thermal populations on the decay rates are negligible (i.e., $\{n_{th,m},\,n_{th,sk}\} \ll 1$), the multimode strong-coupling is still valid.

Additionally, the spatial configuration of the skyrmion may be disturbed by the strain induced by piezoelectric effect~\cite{Skyrmion-Motion-Control3}. To overcome this, the skyrmion should be suspended above the piezoelectric substrate (e.g., by resorting to the flip-chip bonding technology~\cite{SAW-Entanglement1,SAW-Cavity-Multimode-TunableCoupling}) with a distance $d$. Correspondingly, the coupling strength would be reduced to $g_m \rightarrow g_m e^{-k_m d}$. Note that when taking typical distance $d\simeq1\,\SI{}{\micro\meter}$ (approximate to one center wavelength $\lambda_0$), the coupling strength reduce to $g_m \rightarrow g_m e^{-1}$. Even for this, one still ensures the strong coupling between single skyrmion qubit and single phonon of different modes.

%%%%%%%%%%%%%%%%%
\begin{figure}[tb]
	\centering
	\includegraphics[width=8.6cm]{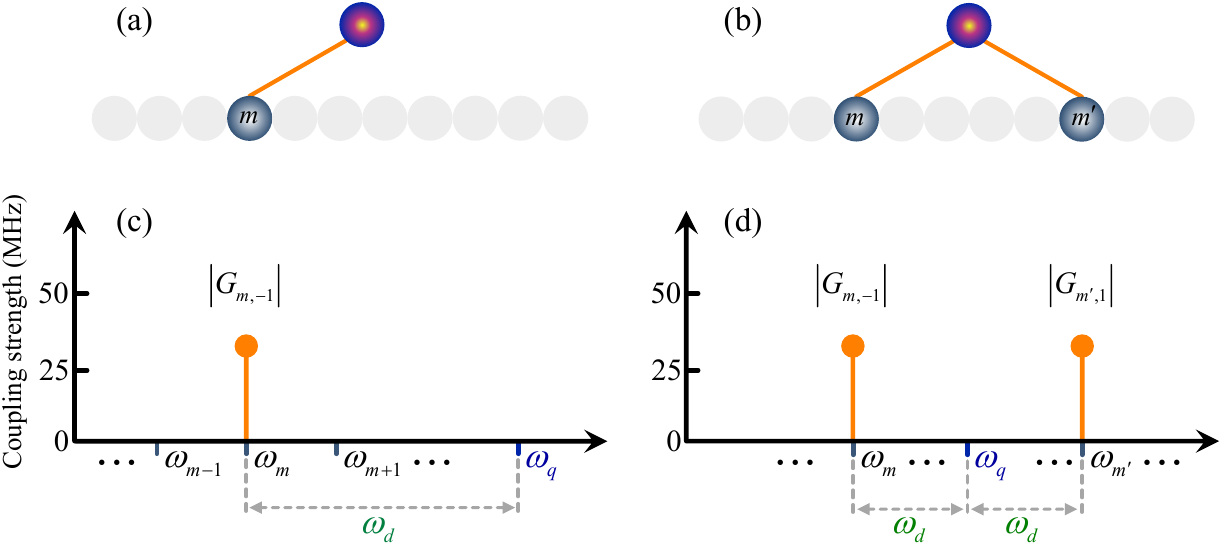}
	\caption{Diagrams for single skyrmion qubit selectively accessing (a) one (b) two of individual SAW phononic modes. Here, the dark (light) gray balls denote the phononic modes that are coupled (uncoupled) to qubit. (c) and (d) Frequency space diagram of phononic modes and qubit for the cases shown in (a) and (b). The parameters are taken as $A_z=0.5\omega_d$, with $\omega_d$ determined by the frequency difference between qubit and target phononic mode. Other parameters are the same as those in Fig.~\ref{F2StrongCoupling}.}
\label{F3AccessModes}
\end{figure}
%%%%%%%%%%%%%%%%%

%%%%%%%%%%%%%%%%%%%%%%%%%%%%%%%%%%%%%%%%%%%%%%%%%%%%%%%%%%%%%%
{\it Skyrmion qubit accessing individual phononic modes.---}
\label{AccessIndividualModes}
When phonons of different modes serving as data bus or registers, the selective accessing of individual phononic modes is an essential operation for initialization and manipulation of data. Equation~(\ref{EQ1Hamiltonian}) shows that we can tune the qubit frequency such that $\omega_{q}=\omega_{m}$, then the qubit can be selective coupled to the individual SAW phononic mode. However, if the frequency of the Skyrmion qubit is fixed, we can use sideband technique to access specific phononic mode, as illustrated in Figs.~\ref{F3AccessModes}(a)~and~~\ref{F3AccessModes}(b). That is, a time-dependent magnetic field~\cite{Skyrmion-Modulation-Field} $h_{zd}^{(l)}=A_z \text{cos}(\omega_d t)$, with amplitude $A_z$ and frequency $\omega_d$, is taken to longitudinally modulate the qubit, with the interaction Hamiltonian given by $A_z \text{cos}(\omega_d t) \sigma_{11}$. Then the effective interaction Hamiltonian between the qubit and the photon modes is given as~\cite{Supplemental-Material}
\begin{align}
H_{QM}&=\sum_{m,n} G_{m,n}
\big[b_m \sigma_{10} e^{i(\omega_q -\omega_m +n\omega_d)t} +\text{H.c.}\big],
\label{EQ3Hamiltonian-Driven}
\end{align}
with the coupling strength $G_{m,n}=g_m J_n$ between the $n$th-order sideband of the qubit and single phonon of mode $m$. $J_n \equiv J_n (A_z/\omega_d)$ is the $n$th-order Bessel function.

It is clear that the qubit can be selectively coupled to phonons of different modes by using the frequency matching~\cite{Qubit-modulation1,Qubit-modulation2} condition $\omega_{q}-\omega_{m}+n\omega_{d}=0$, as indicated in Eq.~(\ref{EQ3Hamiltonian-Driven}), even the frequency of qubit is detuned from phononic frequencies, i.e., $\omega_{q} \neq \omega_{m}$. Note that one can control the coupling strength $G_{m,n}$ by adjusting the amplitude $A_z$ and frequency $\omega_d$. Based on this, when tuning the qubit off-resonance with the target phononic mode (e.g., $\omega_q \neq \omega_m$), by taking the frequency $\omega_d$ to match the frequency difference between qubit and target phononic mode, e.g., $\omega_d =\omega_q - \omega_m$ as shown in Fig.~\ref{F3AccessModes}(c), one ensures the frequency matching condition $\omega_{q} -\omega_{m} +n\omega_{d}=0$ for $n=-1$. That means, the $-1$-th sideband of qubit is resonantly coupled to the phonon of frequency $\omega_m$ with the coupling strength $G_{m,-1}$, while other sidebands (i.e., $n\neq-1$) are detuned from the phononic frequency $\omega_m$. Thus, the qubit can selectively access one of individual phononic modes. Here, the corresponding coupling strength is much larger than the decay rate, i.e., $|G_{m,-1}| \gg \{\Gamma_{sk},\,\gamma_{s,m}\}$, indicating such accessing operates in strong-coupling regime.

Furthermore, when tuning the frequency $\omega_q$ of qubit exactly halfway between two target phononic modes, e.g., $\omega_q =\left(\omega_{m}+\omega_{m'}\right)/2$, one obtains $\omega_{q}-\omega_{m}=-(\omega_{q}-\omega_{m'})$. By taking the modulation frequency $\omega_d$ to match the frequency difference between qubit and any one of the two target phononic modes, i.e., $\omega_d =\omega_q - \omega_m$ or $\omega_d =\omega_{m'} - \omega_q$ as shown in Fig.~\ref{F3AccessModes}(d), one can ensure the frequency matching conditions $\omega_{q} -\omega_{m} +n\omega_{d}=0$ for $n=-1$ and $\omega_{q} -\omega_{m'} +n\omega_{d}=0$ for $n=1$. That means, the $-1$-th ($+1$-th) sideband of qubit is resonantly coupled to the phonon of frequency $\omega_m$ ($\omega_{m'}$) with the coupling strength $G_{m,-1}$ ($G_{m',1}$), while other sidebands (i.e., $|n|\neq1$) are detuned from the phonon of the two phononic frequency $\omega_m$ and $\omega_{m+j}$. Thus, the qubit can selectively access two of individual phononic modes. This should be highly attractive for the parallel operation, due to the reduction in overall time for quantum information precessing.

%%%%%%%%%%%%%%%%%%%%%%%%%%%%%%%%%%%%%%%%%%%%%%%%%%%%%%%%%%%%%%
{\it Phonon-phonon interaction.---}
\label{PhononPhononInteraction}
When phononic modes serving as quantum register units, it is highly desired to control the interaction between different phononic modes. We can use the qubit assisted by the longitudinal magnetic field $h_{zd}^{(l)}$ to achieve selective control of the phonon-phonon interaction between different modes. In the case that all phononic modes are far-detuned from qubit sidebands and the qubit is at ground state, the effective Hamiltonian of phononic modes is written as~\cite{Supplemental-Material}
\begin{align}
H_{MM}&=
\sum_m \chi_m b_m^{\dagger} b_m+
\nonumber\\
&~~~\sum_{m' \neq m}
G_{m,m'} b_m^{\dagger} b_{m'} e^{i[(n'-n)\omega_d +{\omega}_m-{\omega}_{m'}]t}
+\text{H.c.},
\label{EQ4Hamiltonian-Phonon-Phonon}
\end{align}
where $\chi_m=\sum_{n} g_m^2 J_n^2/(\omega_m-\omega_q-n\omega_d)$ is the frequency shift due to the coupling between the phonon of mode $m$ and sidebands of qubit. $G_{m,m'}=\sum_{n}g_m g_{m'} J_n J_{n'} /2(\omega_m-\omega_q-n\omega_d) +\sum_{n'} g_m g_{m'} J_n J_{n'} /2(\omega_{m'}-\omega_q-{n'}\omega_d)$ denotes the coupling strength between two phonons of modes $m$ and ${m'}$.

\begin{figure}[tb]
	\centering
	\includegraphics[width=8.6cm]{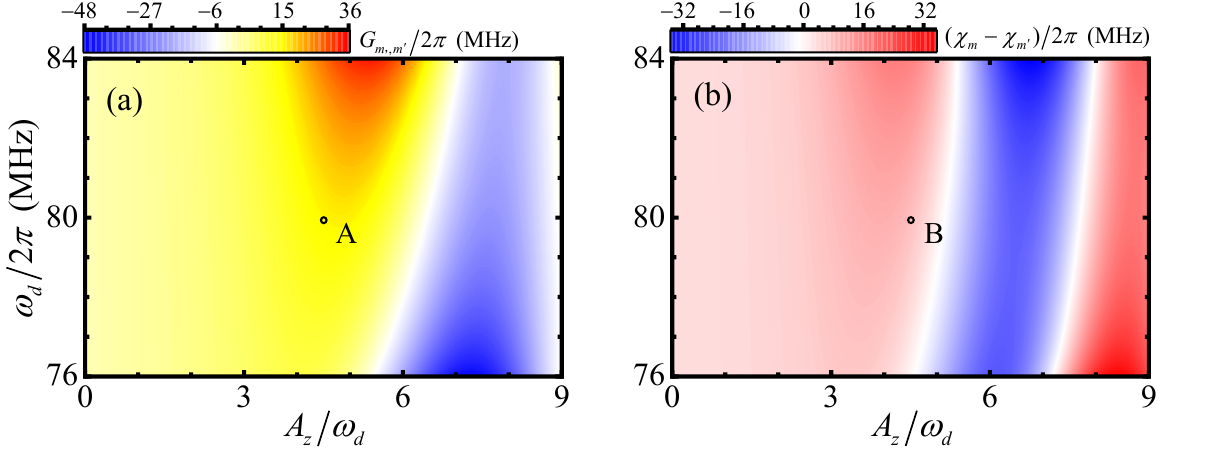}
	\caption{Taking the phonon-phonon interaction between modes $m=-2$ and $m'=2$ as an example, (a) and (b) show the coupling strength $G_{m,m'}$ and frequency shift difference $\chi_m -\chi_{m'}$ between the two modes versus the amplitude $A_z$ and frequency $\omega_d$. Points A and B correspond to the case satisfying $(n-n')\omega_d +(\omega_{m'}+\chi_{m'}) -(\omega_{m}+\chi_{m})=0$. Other parameters are the same as those in Fig.~\ref{F2StrongCoupling} except $(\omega_q -\omega_0)/2\pi=400\,\text{MHz}$.}
\label{F4PhononPhononCoupling}
\end{figure}

As indicated in Eq.~(\ref{EQ4Hamiltonian-Phonon-Phonon}), the phonon-phonon interaction between two different modes $m$ and $m'$ can be activated by adjusting the amplitude $A_z$ and frequency $\omega_d$, while the corresponding coupling strength $G_{m,m'}$ is controlled by adjusting the amplitude $A_z$ and frequency $\omega_d$. Based on this, when tuning $\omega_d$ to match the frequency difference between two target phononic modes. i.e., $\omega_d=\omega_{m'}-\omega_m$, one can first ensures the frequency matching condition $(n'-n)\omega_d +\omega_m -\omega_{m'}=0$ for $n'=n+1$. However, as indicated in Eq.~(\ref{EQ4Hamiltonian-Phonon-Phonon}), the dispersive couplings between phononic modes and qubit sidebands can lead to unique frequency shifts $\chi_m$ for different phononic modes. In order to selectively active the coupling between two phonons of target modes $m$ and $m'$, one should slightly adjust the amplitude $A_z$ and frequency $\omega_d$ to match the modified frequency matching condition $(n'-n)\omega_d +(\omega_{m}+\chi_{m}) -(\omega_{m'}+\chi_{m'})=0$. In Figs.~\ref{F4PhononPhononCoupling}(a) and~\ref{F4PhononPhononCoupling}(b), we take the phonon-phonon interaction between target modes $m=-2$ and $m'=2$ as an example to show the coupling strength $G_{m,m'}$ and frequency shift difference $\chi_m -\chi_{m'}$ versus the amplitude $A_z$ and frequency $\omega_d$. We find that by taking $\omega_d/2\pi = 80\,\text{MHz}$ and $A_z/\omega_d \simeq 5$, one can ensure $(n'-n)\omega_d +(\omega_{-2}+\chi_{-2}) -(\omega_{+2}+\chi_{+2})=0$, enabling selective activation of the phonon-phonon interaction between these two modes. Here, the coupling strength ($G_{m,m'}/2\pi \simeq 10\,\text{MHz}$) is much larger than the decay rates of qubit and phonons. Thus, the qubit-mediated phonon-phonon interaction also operates in strong-coupling regime.

%%%%%%%%%%%%%%%%%
%%%%%%%%%%%%%%%%%

%%%%%%%%%%%%%%%%%
\begin{figure}[tb]
	\centering
	\includegraphics[width=8.6cm]{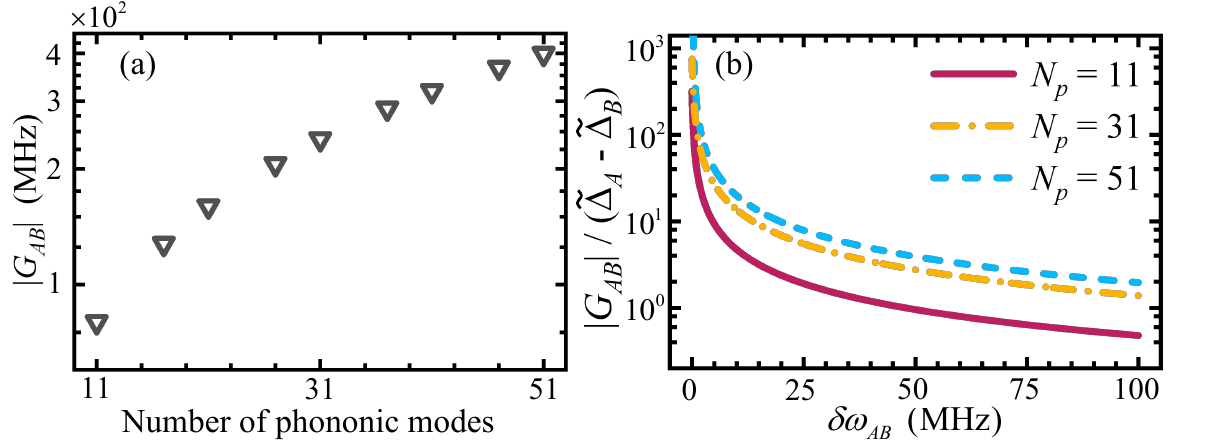}
	\caption{(a) Coupling strengths $|G_{AB}|$ between two qubits for different numbers $N_p$ of phononic modes. (b) Ratio between the coupling strengths $|G_{AB}|$ and effective detuning $|\tilde{\Delta}_A -\tilde{\Delta}_B|$ versus the detuning $\delta\omega_{AB}$ for different numbers of phononic modes. Other parameters are the same as those in Fig.~\ref{F2StrongCoupling} except $x_A=L_c/2 +\lambda_0/4$, $x_B=L_c/2 -7\lambda_0/4$, and $\left[(\omega_A+\omega_B)/2\right]/2\pi=3.2\,\text{GHz}$.}
\label{F5SkyrmionSkyrmionCoupling}
\end{figure}
%%%%%%%%%%%%%%%%%

{\it Skyrmion-skyrmion interaction.---}
\label{QubitQubitInteraction}
We now study the interaction between two skyrmion qubits mediated by phononic modes in a common SAW cavity. In the case that all phononic modes are far-detuned from the qubits and in vacuum state, the effective interaction Hamiltonian between two qubits is given as~\cite{Supplemental-Material}
\begin{align}
H_{QQ}&=\tilde{\Delta}_A \sigma_{11}^A+ \tilde{\Delta}_B \sigma_{11}^B
+G_{AB} \left(\sigma_{01}^A\sigma_{10}^B +\sigma_{10}^A\sigma_{01}^B\right).
\label{EQ5Hamiltonian-Qubit-Qubit}
\end{align}
$\tilde{\Delta}_{A}=(\omega_A-\omega_B)/2-\sum_m {(g_m^{A})^2 }\big/{\Delta_m}$ (or $\tilde{\Delta}_{B}=(\omega_B-\omega_A)/2-\sum_m {(g_m^{B})^2}\big/{\Delta_m}$) is the effective detuning between qubit A (or qubit B) with frequency $\omega_A$ (or $\omega_B$) and their average transition frequency $(\omega_A+\omega_B)/2$, shifted by phonons of different modes. $G_{AB}=-\sum_m {g_m^A g_m^B}\big/{\Delta_m}$ is the coupling strength between two qubits mediated by all phononic modes with detuning $\Delta_m=\omega_m-(\omega_A+\omega_B)/2$.

As shown in Eq.~(\ref{EQ5Hamiltonian-Qubit-Qubit}), the effective coupling strength $G_{AB}$ between two qubits highly depends on the number of phononic modes. Figure~\ref{F5SkyrmionSkyrmionCoupling}(a) shows the dependence of $|G_{AB}|$ on the number $N_p$ of phononic modes. Here, the two qubits are taken to be located at different positions $x_A=L_c/2 +\lambda_0/4$ and $x_B=L_c/2 -7\lambda_0/4$, respectively. We find that the increase of $N_p$ leads to the enhancement of $|G_{AB}|$, since more phononic modes are introduced to mediate the coupling between qubits. Moreover, the interaction between qubits suffers from the detuning $\delta\omega_{AB}=|\omega_A-\omega_B|$ between the bare transition frequencies of two qubits, and such a detuning would be modified by phonons as indicated in Eq.~(\ref{EQ5Hamiltonian-Qubit-Qubit}). To ensure the strong interaction between qubits, the coupling strength $G_{AB}$ between qubits should be much larger than the effective detuning~$|\widetilde{\Delta}_A-\widetilde{\Delta}_B|$. As shown in Fig.~\ref{F5SkyrmionSkyrmionCoupling}(b), with the increase of $N_p$, $|G_{AB}/(\widetilde{\Delta}_A-\widetilde{\Delta}_B)|$ would increase accordingly. That means, the increase of the phononic mode number helps to decrease the effect of dutuning $\delta\omega_{AB}$ on the interaction between qubits. Here, the coupling strength between qubits is much larger than the decay rate of single qubit, i.e., $|G_{AB}| \gg \Gamma_{sk}$. Thus, the phonons-mediated qubit-qubit interaction operates in strong-coupling regime.

%%%%%%%%%%%%%%%%%%%%%%%%%%%%%%%%%%%%%%%%%%%%%%%%%%%%%%%%%%%%%%
{\it Experimental feasibility.---}
\label{Feasibility}
Skyrmions have been experimentally studied~\cite{Skyrmion-Review1,Skyrmion-Review2,Skyrmion-Review-Devices1,Skyrmion-Review-Devices2,Skyrmion-Review-Devices3} in various magnetic materials, e.g., NiGa$_2$S$_4$~\cite{Skyrmion-Material-NiGa2S4}, $\alpha$-NaFeO$_2$~\cite{Skyrmion-Material-aNaFeO2}, Fe$_x$Ni$_{1-x}$Br$_2$~\cite{Skyrmion-Material-FexNi1xBr2}, Gd$_2$PdSi$_3$~\cite{Skyrmion-Material-Gd2PdSi3}, and Gd$_3$Ru$_4$Al$_2$~\cite{Skyrmion-Material-Gd3Ru4Al2}. Here, the skyrmion qubit is considered to be generated in a magnetic film, with the thickness much smaller than the wavelength of SAW. The parameters of skyrmion are taken as~\cite{SkyrmionQubit-First-Theory,SkyrmionQubit-QuantumComputing,SkyrmionQubit-Magnon}: effective spin $\overline{S}=10$, lattice spacing $a=0.5\,\text{nm}$, interaction strength $J_1=3\,\text{meV}$, anisotropy coupling $K_z=0.3\,\text{meV}$, electric polarization $P_E=0.2$\,C/m, and static magnetic field $H_z\sim100\,\text{mT}$. The thin film allows for a low decay rate $\Gamma_{sk}/2\pi \simeq 0.4\,\text{MHz}$ of skyrmion. The distance between the centers of two skyrmions is taken as $|x_A-x_B|\simeq 2\,\SI{}{\micro\meter}$, much larger than the diameter of single skyrmion. This ensures the skyrmions are spatially isolated and would not suffer from attraction or repulsion interaction. The position adjustment of skyrmions requires the stabilization and individual manipulation of skyrmions, which have been realized in experiment and applied in skyrmion-based logical device and racetrack memory~\cite{Skyrmion-Review-Devices1,Skyrmion-Review-Devices2,Skyrmion-Review-Devices3,Skyrmion-Review1,Skyrmion-Motion-Control1,
Skyrmion-Motion-Control2}. The generation and manipulation of an array of nanoscale skyrmions have been realized in experiments~\cite{Skyrmion-Array1,Skyrmion-Array2}, which allows coupling many skyrmion qubits to phonons of different modes in single SAW cavity.

With rapid development of microfabrication lithography technology, SAW cavities, which are deposited on substrate made of piezoelectric materials (e.g., Quartz~\cite{Piezoelectric-Material-SiO2}, GaAs~\cite{Piezoelectric-Material-GaAs}, Bi$_{12}$GeO$_{20}$~\cite{Piezoelectric-Material-Bi12GeO20}, LiNbO$_3$~\cite{Piezoelectric-Material-LiNbO3}, and ZnO~\cite{Piezoelectric-Material-ZnO}) and support long-lived phonons of different modes, have been realized experimentally~\cite{SAW-Background-Book1,SAW-Background-Book2,SAW-Cavity-Review,SAW-Cavity-Multimode-TunableCoupling}. Here, we take the piezoelectric material $128^{\circ}$ Y-Z cut LiNbO$_3$~\cite{QuantumAcoustics-Transducer1-Article1-110-PESAW-Quantization,SAW-Cavity-Multimode-TunableCoupling} as substrate and choose Rayleigh-type SAW that has out-of-plane mechanical component. In this case, the propagation velocity of SAW is $v_s=3979$\,m/s, and the skyrmion qubit is coupled to SAW phonons via out-of-plane electric field induced by piezoelectric effect. To obtain a SAW cavity with the center frequency, e.g., $\omega_0/2\pi=4\,\text{GHz}$ corresponding to the center wavelength $\lambda_0=v_s/\omega_0\simeq1\,\SI{}{\micro\meter}$, the period of each Bragg grating is taken as $\Lambda_g=\lambda_0/2$. To ensure the SAW cavity support multiple phononic modes, the effective length of cavity, electrode number of Bragg grating, and reflectivity of each grating electrode can be designed as, e.g., $L_c \sim 100\lambda_0$, $N_g=200$, and $r_g=0.01$, such that the free spectral range is smaller than the reflection bandwidth of grating. Moreover, the width of each Bragg grating and the number of IDT period are designed as $W=100\,\SI{}{\micro\meter}$ and $N_i=10$ to ensure low decay rates of phonons, while the IDT period is taken as $\Lambda_i=\lambda_0$.

%%%%%%%%%%%%%%%%%%%%%%%%%%%%%%%%%%%%%%%%%%%%%%%%%%%%%%%%%%%%%%
{\it Conclusion.---}
\label{Conclusion}
We have proposed and studied a hybrid quantum system composed of skyrmion qubit and a SAW cavity supporting a number of phononic modes. The interaction between single qubit and single phonon of different modes can operate in strong-coupling regime. By manipulating the qubit via static and time-dependent modulation magnetic fields, one can achieve the single qubit selectively accessing specific phononic modes, phonon-phonon interaction between two specific modes, and qubit-qubit interaction mediated by phonons of different modes, which also operate in strong-coupling regime. Owing to the controllability of nanoscale skyrmion qubit and the dense long-lived phononic modes in single SAW cavity at scale of hundreds of micrometers, our system can be a promising candidate for miniaturized hybrid microwave devices. Such devices integrate many skyrmion qubits (as storage units) with single SAW cavity (as data bus or register), offering potential applications in on-chip quantum information processing.

%%%%%%%%%%%%%%%%%%%%%%%%%%%%%%%%%%%%%%%%%%%%%%%%%%%%%%%%%%%%%%
{\it Acknowledgements.}
This work was supported by the National Natural Science Foundation of China Grants No. 12374483, No. 92365209, and No. 62474012.


\begin{thebibliography}{99}
\bibitem{Superconducting-Circuit1}Z.-L. Xiang, S. Ashhab, J. Q. You, and F. Nori, Hybrid quantum circuits: Superconducting circuits interacting with other quantum systems, Rev. Mod. Phys.~\textbf{85}, 623 (2013).
	
\bibitem{Superconducting-Circuit2}X. Gu, A. F. Kockum, A. Miranowicz, Y.-X. Liu, and F. Nori, Microwave photonics with superconducting quantum circuits, Phys. Rep.~\textbf{718-719}, 1 (2017).

\bibitem{Superconducting-Circuit3}M. Kjaergaard, M. E. Schwartz, J. Braum\"{u}ller, P. Krantz, J. I. J. Wang, S. Gustavsson, and W. D. Oliver, Superconducting qubits: Current state of play, Annu. Rev. Conden. Ma. P.~\textbf{11}, 369 (2020).



\bibitem{Superconducting-Hybrid}A. A. Clerk, K. W. Lehnert, P. Bertet, J. R. Petta, and Y. Nakamura, Hybrid quantum systems with circuit quantum electrodynamics, Nat. Phys. \textbf{16}, 257 (2020).

\bibitem{Superconducting-Hybrid-Chip}X.-B. Xu, W.-T. Wang, L.-Y. Sun, and C.-L. Zou, Hybrid superconducting photonic-phononic chip for quantum information processing, Chip \textbf{1}, 100016 (2022).



\bibitem{Cavity-Bus1}M. A. Sillanp\"{a}\"{a}, J. I. Park, and R. W. Simmonds, Coherent quantum state storage and transfer between two phase qubits via a resonant cavity, Nature \textbf{449}, 438 (2007).

\bibitem{Cavity-Bus2}J. Majer, J. M. Chow, J. M. Gambetta, Jens Koch, B. R. Johnson, J. A. Schreier, L. Frunzio, D. I. Schuster, A. A. Houck, A. Wallraff, A. Blais, M. H. Devoret, S. M. Girvin, and R. J. Schoelkopf, Coupling superconducting qubits via a cavity bus, Nature \textbf{449}, 443 (2007).

\bibitem{Cavity-Register1}F. Altomare, J. I. Park, K. Cicak, M. A. Sillanp\"{a}\"{a}, M. S. Allman, D. Li, A. Sirois, J. A. Strong, J. D. Whittaker, and R. W. Simmonds, Tripartite interactions between two phase qubits and a resonant cavity, Nat. Phys. \textbf{6}, 777 (2010).

\bibitem{Cavity-Register2}M. Mariantoni, H. Wang, R. C. Bialczak, M. Lenander, E. Lucero, M. Neeley, A. D. O' Connell, D. Sank, M. Weides, J. Wenner, T. Yamamoto, Y. Yin, J. Zhao, J. M. Martinis, and A. N. Cleland, Photon shell game in three-resonator circuit quantum electrodynamics, Nat. Phys. \textbf{7}, 287 (2011).



\bibitem{QuantumAcoustics-SAW-Superconducting-Coupling1}M. V. Gustafsson, T. Aref, A. F. Kockum, M. K. Ekstr\"{o}m, G. Johansson, and P. Delsing, Propagating phonons coupled to an artificial atom, Science \textbf{346}, 207 (2014).

\bibitem{QuantumAcoustics-Transducer1-Article1-110-PESAW-Quantization}M. J. A. Schuetz, E. M. Kessler, G. Giedke, L. M. K. Vandersypen, M. D. Lukin, and J. I. Cirac, Universal quantum transducers based on surface acoustic waves, Phys. Rev. X \textbf{3}, 031031 (2015).

\bibitem{QuantumAcoustics-Article2}R. Manenti, A. F. Kockum, A. Patterson, T. Behrle, J. Rahamim, G. Tancredi, F. Nori, and P. J. Leek, Circuit quantum acoustodynamics with surface acoustic waves, Nat. Commun. \textbf{8}, 975 (2017).

\bibitem{SAW-Sensing}A. Noguchi, R. Yamazaki, Y. Tabuchi, and Y. Nakamura, Qubit-assisted transduction for a detection of surface acoustic waves near the quantum limit, Phys. Rev. Lett. \textbf{119}, 180505 (2017).

\bibitem{QuantumAcoustics-Multimode-AW}C. T. Hann, C. L. Zou, Y. X. Zhang, Y. W. Chu, R. J. Schoelkopf, S. M. Girvin, and L. Jiang, Hardware-efficient quantum random access memory with hybrid quantum acoustic systems, Phys. Rev. Lett. \textbf{123}, 250501 (2019).

\bibitem{QuantumAcoustics-Waveguide}G. Andersson, B. Suri, L. Guo, T. Aref, and P. Delsing, Non-exponential decay of a giant artificial atom, Nat. Phys. \textbf{15}, 1123 (2019).

\bibitem{SAW-Entanglement1}A. Bienfait, K. J. Satzinger, Y. P. Zhong, H.-S. Chang, M.-H. Chou, C. R. Conner, \'{E}. Dumur, J. Grebel, G. A. Peairs, R. G. Povey, and A. N. Cleland, Phonon-mediated quantum state transfer and remote qubit entanglement, Science \textbf{364}, 368 (2019).

\bibitem{QuantumAcoustics-SAW-Superconducting-Coupling2}G.-H. Zeng, Y. Zhang, A. N. Bolgar, D. He, B. Li, X.-H. Ruan, L. Zhou, L. M. Kuang, O. V. Astafiev, Y.-X. Liu, and Z. H. Peng, Quantum versus classical regime in circuit quantum acoustodynamics, New J. Phys. \textbf{23}, 123001 (2021).

\bibitem{SAW-Entanglement2}G. Andersson, S. W. Jolin, M. Scigliuzzo, R. Borgani, M. O. Thol\'{e}n, J. C. R. Hern\'{a}ndez, V. Shumeiko, D. B. Haviland, and P. Delsing, Squeezing and multimode entanglement of surface acoustic wave phonons, PRX Quantum \textbf{3}, 010312 (2022).

\bibitem{SAW-Piezomagnetism}Y.-Y. Chen, J.-H. Wang, L. N. Song, and Y.-X. Liu, Manipulation of magnetic systems by quantized surface acoustic wave via piezomagnetic effect, Phys. Rev. Applied \textbf{23}, 034013 (2025).



    	
\bibitem{SAW-Background-Book1}S. Datta, \textit{Surface Acoustic Wave Devices} (Prentice-Hall, Upper Saddle River, NJ, 1986).
	
\bibitem{SAW-Background-Book2}D. Morgan, \textit{Surface Acoustic Wave Filters} (Academic Press, Boston, 2007).


\bibitem{SAW-Sensor}D. Mandal and S. Banerjee, Surface acoustic wave (SAW) sensors: physics, materials, and applications, Sensors \textbf{22}, 820 (2022).

\bibitem{SAW-Fluid-Control}L. Y. Yeo and J. R. Friend, Surface acoustic wave microfluidics, Annu. Rev. Fluid Mech. \textbf{46}, 379 (2014).



\bibitem{SAW-Cavity}W. H. Haydl, B. Dischler, and P. Hiesinger, Multimode SAW resonators: a method to study the optimum resonator design, Proc. IEEE Ultrason. Symp. \textbf{1}, 287 (1976).

\bibitem{QuantumAcoustics-Multimode-SAW}B. A. Moores, L. R. Sletten, J. J. Viennot, and K. W. Lehnert, Cavity quantum acoustic device in the multimode strong coupling regime, Phys. Rev. Lett. \textbf{120}, 227701 (2018).

\bibitem{SAW-Cavity-Multimode-TunableCoupling}X. H. Ruan, L. Li, G. Liang, S. Zhao, J. Wang, Y. Bu, B. Chen, X. Song, X. Li, H. Zhang, J. Wang, Q. Zhao, K. Xu, H. Fan, Y. X. Liu, J. Zhang, Z. H. Peng, Z. C. Xiang, and D. N. Zheng, Tunable coupling of a quantum phononic resonator to a transmon qubit via galvanic-contact flip-chip architecture, Appl. Phys. Lett.~\textbf{125}, 052603 (2024).



\bibitem{Skyrmion-First-Theory}A. N. Bogdanov and U. K. R\"{o}$\beta$ler, Chiral symmetry breaking in magnetic thin films and multilayers, Phys. Rev. Lett. \textbf{87}, 037203 (2001).

\bibitem{Skyrmion-Review-Devices1}G. Finocchio, F. B\"{u}ttner, R. Tomasello, M. Carpentieri, and M. Kl\"{a}ui, Magnetic skyrmions: from fundamental to applications, J. Phys. D \textbf{49}, 423001 (2016).

\bibitem{Skyrmion-Review-Devices2}A. Fert, N. Reyren, and V. Cros, Magnetic skyrmions: advances in physics and potential applications, Nat. Rev. Mater. \textbf{2}, 17031 (2017).

\bibitem{Skyrmion-Review-Devices3}X. Zhang, Y. Zhou, K. M. Song, T. E. Park, J. Xia, M. Ezawa, X. Liu, W. Zhao, G. Zhao, and S. Woo, Skyrmion-electronics: writing, deleting, reading and processing magnetic skyrmions toward spintronic applications, J. Phys. Condens. Matter \textbf{32}, 143001 (2020).

\bibitem{Skyrmion-Review1}C. Reichhardt, C. J. O. Reichhardt, and M. V. Milo\v{s}evi\'{c}, Statics and dynamics of skyrmions interacting with disorder and nanostructures, Rev. Mod. Phys. \textbf{94}, 035005 (2022).

\bibitem{Skyrmion-Review2}A. P. Petrovi\'{c}, C. Psaroudaki, P. Fischer, M. Garst, and C. Panagopoulos, Colloquium: quantum properties and functionalities of magnetic skyrmions, arXiv:~2410.11427~(2024).




\bibitem{SkyrmionQubit-First-Theory}C. Psaroudaki and C. Panagopoulos, Skyrmion qubits: a new class of quantum logic elements based on nanoscale magnetization, Phys. Rev. Lett. \textbf{127}, 067201 (2021).
    
\bibitem{SkyrmionQubit-First-Theory-Helicity}C. Psaroudaki and C. Panagopoulos, Skyrmion helicity: Quantization and quantum tunneling effects, Phys. Rev. B \textbf{106}, 104422 (2022).

\bibitem{SkyrmionQubit-First-Review}C. Psaroudaki, E. Peraticos, and C. Panagopoulos, Skyrmion qubits: challenges for future quantum computing applications, Appl. Phys. Lett. \textbf{123}, 260501 (2023).

\bibitem{SkyrmionQubit-QuantumComputing}J. Xia, X. C. Zhang, X. X. Liu, Y. Zhou, and M. Ezawa, Universal quantum computation based on nanoscale skyrmion helicity qubits in frustrated magnets, Phys. Rev. Lett. \textbf{130}, 106701 (2023).

\bibitem{SkyrmionQubit-Magnon}X. F. Pan, P. B. Li, X. L. Hei, X. C. Zhang, M. Mochizuki, F. L. Li, and F. Nori, Magnon-Skyrmion hybrid quantum systems: tailoring interactions via magnons, Phys. Rev. Lett. \textbf{132}, 193601 (2024).


\bibitem{MagneticSpins-Electric-Control}H. Katsura, N. Nagaosa, and A. V. Balatsky, Spin current and magnetoelectric effect in noncollinear magnets, Phys. Rev. Lett. \textbf{95}, 057205 (2005).
    
\bibitem{Skyrmion-Competing-Interactions}S.-Z. Lin and S. Hayami, Ginzburg-Landau theory for skyrmions in inversion-symmetric magnets with competing interactions, Phys. Rev. B \textbf{93}, 064430 (2016).

\bibitem{Supplemental-Material}See Supplemental Material at http://link.aps.org/... for detail derivations.



\bibitem{Skyrmion-Motion-Control1}S. Woo, K. Litzius, B. Kr\"{u}ger, M. Y. Im, L. Caretta, K. Richter, M. Mann, A. Krone, R. M. Reeve, M. Weigand, P. Agrawal, I. Lemesh, M. A. Mawass, P. Fischer, M. Kl\"{a}ui, and G. S. D. Beach, Observation of room-temperature magnetic skyrmions and their current-driven dynamics in ultrathin metallic ferromagnets, Nat. Mater. \textbf{15}, 501 (2016).

\bibitem{Skyrmion-Motion-Control2}S. L. Zhang, W. W. Wang, D. M. Burn, H. Peng, H. Berger, A. Bauer, C. Pfleiderer, G. van der Laan, and T. Hesjedal, Manipulation of skyrmion motion by magnetic field gradients, Nat. Commun. \textbf{9}, 2115 (2018).

\bibitem{Skyrmion-Motion-Control3}T. Yokouchi, S. Sugimoto, B. Rana, S. Seki, N. Ogawa, S. Kasai, and Y. Otani, Creation of magnetic skyrmions by surface acoustic waves, Nat. Nanotechnol. \textbf{15}, 361 (2020).



\bibitem{Skyrmion-Modulation-Field}W. W. Wang, M. Beg, B. Zhang, W. Kuch, and H. Fangohr, Driving magnetic skyrmions with microwave fields, Phys. Rev. B \textbf{92}, 020403(R) (2015).



\bibitem{Qubit-modulation1}Y.-X. Liu, L. F. Wei, J. R. Johansson, J. S. Tsai, and F. Nori, Superconducting qubits can be coupled and addressed as trapped ions, Phys. Rev. B \textbf{76}, 144518 (2007).
    
\bibitem{Qubit-modulation2}Y.-X. Liu, C.-X. Yang, H.-C. Sun, and X.-B. Wang, Coexistence of single- and multi-photon processes due to longitudinal couplings between superconducting flux qubits and external fields, New J. Phys. \textbf{16}, 015031 (2014).



\bibitem{Skyrmion-Material-NiGa2S4}S. Nakatsuji, Y. Nambu, H. Tonomura, O. Sakai, S. Jonas, C. Broholm, H. Tsunetsugu, Y. M. Qiu, and Y. Maeno, Spin disorder on a triangular lattice, Science \textbf{309}, 1697 (2005).

\bibitem{Skyrmion-Material-aNaFeO2}T. McQueen, Q. Huang, J. W. Lynn, R. F. Berger, T. Klimczuk, B. G. Ueland, P. Schiffer, and R. J. Cava, Magnetic structure and properties of the $S=5/2$ triangular antiferromagnet $\alpha$-NaFeO$_2$, Phys. Rev. B \textbf{76}, 024420 (2007).

\bibitem{Skyrmion-Material-FexNi1xBr2}M. W. Moore and P. Day, Magnetic phase diagrams and helical magnetic phases in $M_x$Ni$_{1-x}$Br$_2$ ($M =$ Fe, Mn): A neutron diffraction and magneto-optical study, J. Solid State Chem. \textbf{59}, 23 (1985).

\bibitem{Skyrmion-Material-Gd2PdSi3}T. Kurumaji, T. Nakajima, M. Hirschberger, A. Kikkawa, Y. Yamasaki, H. Sagayama, H. Nakao, Y. Taguchi, T.-h. Arima, and Y. Tokura, Skyrmion lattice with a giant topological Hall effect in a frustrated triangular-lattice magnet, Science \textbf{365}, 914 (2019).

\bibitem{Skyrmion-Material-Gd3Ru4Al2}M. Hirschberger, T. Nakajima, S. Gao, L. Peng, A. Kikkawa, T. Kurumaji, M. Kriener, Y. Yamasaki, H. Sagayama, H. Nakao, K. Ohishi, K. Kakurai, Y. Taguchi, X. Yu, T.-h. Arima, and Y. Tokura, Skyrmion phase and competing magnetic orders on a breathing kagom\'{e} lattice, Nat. Commun. \textbf{10}, 5831 (2019).



\bibitem{Skyrmion-Array1}J. Iwasaki, M. Mochizuki, and N. Nagaosa, Universal current-velocity relation of skyrmion motion in chiral magnets, Nat. Commun. \textbf{4}, 1463 (2013).

\bibitem{Skyrmion-Array2}P. Tengdin, B. Truc, A. Sapozhnik, L. Kong, N. del Ser, S. Gargiulo, I. Madan, T. Sch\"{o}nenberger, P. R. Baral, P. Che, A. Magrez, D. Grundler, H. M. R\o nnow, T. Lagrange, J. D. Zang, A. Rosch, and F. Carbone, Imaging the ultrafast coherent control of a skyrmion crystal, Phys. Rev. X \textbf{12}, 041030 (2022).



\bibitem{Piezoelectric-Material-SiO2}R. Bechmann, Elastic and piezoelectric constants of alpha-quartz, Phys. Rev. \textbf{110}, 1060 (1958).

\bibitem{Piezoelectric-Material-GaAs}J. J. Campbell and W. R. Jones, Propagation of piezoelectric surface waves on cubic and hexagonal crystals, J. Appl. Phys. \textbf{41}, 2796 (1970).

\bibitem{Piezoelectric-Material-Bi12GeO20}A. J. Slobodnik and J. C. Sethares, Elastic, piezoelectric, and dielectric constants of Bi$_{12}$GeO$_{20}$, J. Appl. Phys. \textbf{43}, 247 (1972).

\bibitem{Piezoelectric-Material-LiNbO3}K. Wade and A. J. Banister, \textit{The chemistry of aluminium, gallium, indium and thallium} (Elmsford, NY, 1975).

\bibitem{Piezoelectric-Material-ZnO}S. Fujishima, Piezoelectric devices for frequency control and selection in Japan, IEEE Ultrason. Symp. Proc. \textbf{1}, 87 (1990).

\bibitem{SAW-Cavity-Review}C. K. Kent, N. Ramakrishnan, and H. P. Kesuma, Advancements in one-port surface acoustic wave (SAW) resonators for sensing applications: A review, IEEE Sens. J. \textbf{24}, 17337 (2024).



\end{thebibliography}
\end{document}